# Transform-Based Feature Map Compression for CNN Inference

Yubo Shi*, Meiqi Wang*, Siyi Chen*, Jinghe Wei[†] and Zhongfeng Wang*
Email: {mf1923134, mqwang, mg1923066}@smail.nju.edu.cn   pume1975_cnjs@sina.com   zfwang@nju.edu.cn
*School of Electronic Science and Engineering, Nanjing University, China
[†]China Key System & Integrated Circuit Co. Ltd, China

*Abstract*—To achieve higher accuracy in machine learning tasks, very deep convolutional neural networks (CNNs) are designed recently. However, the large memory access of deep CNNs will lead to high power consumption. A variety of hardware-friendly compression methods have been proposed to reduce the data transfer bandwidth by exploiting the sparsity of feature maps. Most of them focus on designing a specialized encoding format to increase the compression ratio. Differently, we observe and exploit the sparsity distinction between activations in earlier and later layers to improve the compression ratio. We propose a novel hardware-friendly transform-based method named 1D-Discrete Cosine Transform on Channel dimension with Masks (DCT-CM), which intelligently combines DCT, masks, and a coding format to compress activations. The proposed algorithm achieves an average compression ratio of 2.9× (53% higher than the state-of-the-art transform-based feature map compression works) during inference on ResNet-50 with an 8-bit quantization scheme.

## I. INTRODUCTION

Convolutional neural networks have been widely applied in a variety of domains, which achieve great performance in many tasks including image classification [1], object detection [2], and natural language processing [3]. With the evolution of the CNN models [4, 5, 6], more complicated structures with a larger number of parameters promise an improvement of predicting accuracy. However, the high power consumption and the unacceptable latency caused by deeper CNN models hinder their implementation of the mobile devices and Internet-of-Thing (IoT) applications on Field Programmable Gate Array (FPGA) or Application Specific Integrated Circuit (ASIC) with low hardware budget.

To solve this problem, researchers usually try to reduce the computational cost of models by pruning convolutional filters [7], such as Neural Architecture Search (NAS) [8, 9]. However, data transfer is another factor that brings high power consumption and latency in the CNN accelerator. When deep CNN models are deployed on FPGA, all feature map data and filters cannot be loaded on-chip at the same time. As a result, the large data of feature maps on CNN have to be repeatedly transferred between on-chip and off-chip memory, which greatly increases the power consumption and the data transfer bandwidth. Therefore, it is a topic worth exploring to reduce this power consumption and latency by compressing feature maps to improve the performance of CNN models on specialized hardware [10].

Many CNN feature map compression algorithms that use the sparsity of activation have been proposed over the last few years. [11-15] focus on designing hardware-friendly compression encoders and decompression decoders or reducing the calculation cost of sparse data by using a special storage format. [16-19] design algorithms to compress data transferred between GPU and CPU memory during training.

In this paper, we design transform-based methods that improve the sparsity of activations before data are sent into sparsity-based compression modules and further increase the compression ratio for CNN inference.

For fused-layer and quantized CNN models [20] deployed on hardware accelerator, we propose different methods for activations to deeply exploit and improve feature map sparsity, including:

- We observe the specific 2-D frequency domain information of activations, which demonstrates the fact that we can use 2D-DCT to compress feature maps in earlier layers of CNN. Then we combine Approximate Sparsity Preprocessing (ASP) and low-bit quantization to compress feature maps in the later layers, which achieves a compression ratio of 2.6× (+30% over [16]).

- By exploiting the difference in the sparsity of activations in different layers, we propose DCT-CM to compress each activation transferred, which achieves a 53% higher compression ratio than [16].

The basic transforms used in proposed methods are proved hardware-friendly in [21-24].

## II. RELATED WORK

### A. Coding Format and Compression Encoders

Activations extracted by ReLU provide a known sparsity [10], thus several sparsity-based coding formats are designed to compress feature maps: Coordinate (COO) [25], Bitmap [26], Run-Length Coding [27] etc.. Several methods based on these coding formats describe their hardware architecture to reduce computational cost: Cnvlutin [11], SCNN [12], Eyeriss [13], EIE [14], etc.. Cambricon-SE [28] tries to use Huffman and LZW to improve the compression ratio of feature maps.

However, the sparsity of feature maps is dynamic, which limits the compression ratio of prior methods. A few more methods change CNN`s structure to compress feature maps but they introduce much accuracy loss [18, 29]. Besides, when compression algorithms are implemented, non-dictionary is regarded as a symbol of hardware-friendly by the majority of researches [10].

Among the non-dictionary compression works, Zero-Value Compression (ZVC) is a simple and highly effective approach based on frequent-value compression that uses Bitmap to store the compressed data. It can achieve a compression ratio of 2.5× for 60% sparse data. Thus, ZVC is widely-used in different compression-based accelerators: [18, 30].

## B. JPEG for ACTivations (JPEG-ACT)

JPEG [31] is an image compression based on the transformation of the frequency domain by using DCT to save important low-frequency components. JPEG-ACT [16] tries to use the same transformation and similar coding format to compress feature maps during training. It introduces trainable quantization matrixes to compress the wide variety of frequency domain information of feature maps on CNN. JPEG-ACT achieves a higher training accuracy and a higher compression ratio among the training accelerators by compressing activations [17-19]. However, in an inference accelerator, this method will bring extra transfer and computational cost.

For this work, we introduce the hardware-friendly DCT and ZVC, then we fuse them with the mask method which stores a few additional data to achieve a higher compression ratio and a lower hardware cost at the same time.

## III. COMPRESSION ALGORITHM

To compress feature maps for inference, we propose the following transform-based methods to exploit data sparsity in fused-layer CNN [20]. Then we use a sparsity-based compressor, ZVC, to complete the compression procedure. Because the compression algorithm will be deployed on a hardware accelerator, the hardware-friendly and widely-used DCT algorithm becomes our choice.

### A. 2D-Discrete Cosine Transform on Earlier Activations

2D-DCT is a common algorithm to process figures in frequency-domain. Inspired by JPEG compressor, we compress feature maps by dicing them into 8×8 patches, applying 2D-DCT (1), where $\mathbf{X}$ for original data and $\mathbf{Y}$ for transformed data, and then quantizing and compressing. After transfer, data need decompression and IDCT (Inverse-DCT) (2). However, prior work indicates that feature maps have different frequency information from visual images [32]. Therefore we have to deal with the frequency domain information of special dimension carefully, instead of cutting high-frequency components blindly.

Because the frequency domain entropy of earlier activations which always require more storage is lower [16], 2D-DCT with quantization is effective to compress them. Then we observe that the later activations are so sparse (Fig. 1) that lower-bit quantization brings a high enough compression ratio. By combining these two ways as Fig. 2, the compression ratio of feature maps can be improved with the validation accuracy kept for image classification tasks.

$$\mathbf{Y} = \mathbf{A}^T \mathbf{X} \mathbf{A} \quad (1)$$

$$\mathbf{X} = \mathbf{A}^T \mathbf{Y} \mathbf{A} \quad (2)$$

$$\mathbf{A}_{ij} = c(i)\cos(\frac{(j+0.5)\pi}{N}i),$$
$$\text{where } c(i) = \begin{cases} \sqrt{1/N} & i = 0 \\ \sqrt{2/N} & i \neq 0 \end{cases}, N = 8 \quad (3)$$

When deploying this method on accelerators, the encoder and decoder modules should transfer activations to registers in shape 1×8×8 to achieve higher throughput. Besides, it is not recommended to read in data from the same channel repeatedly. E.g., it is proper that convolutional filters slide in the channel dimension of input.

### B. 1D-Discrete Cosine Transform on Channel Dimension with Masks

The 2D-DCT in part A exploits the 2-D frequency domain distinction between different layers to compress feature maps, but the reusability of the transform module is poor. To solve this problem, we propose DCT-CM, which uses the potential redundancy in the channel dimension of feature maps. DCT-CM transforms the feature map on the channel dimension and compresses it by dropping the high-frequency component.

The first step of the DCT-CM is using 1D-DCT on the channel dimension of each 8×1×1 patch of feature maps as (4), where $\mathbf{A}$ is the same as (3). Then, DCT-CM uses masks changing with the convolutional block on the transformed patches, and the shape of the mask is the same as patches, 8×1×1. Masks with different valid regions will screen the transformed feature map and only save the low-frequency component as Fig. 3.

$$\mathbf{Y} = \mathbf{A}\mathbf{X} \quad (4)$$

$$\mathbf{X} = \mathbf{A}^T \mathbf{Y} \quad (5)$$

Although there is redundancy [33] and vacancy in feature maps, 1D-DCT cannot directly exploit them to compress feature maps. We observe that the distribution of values in the channel dimension is random, which implies that the frequency domain entropy is high. We have to reorder the data after transformation by an extra mask, so that the local channels of the model will be low-frequency during training,

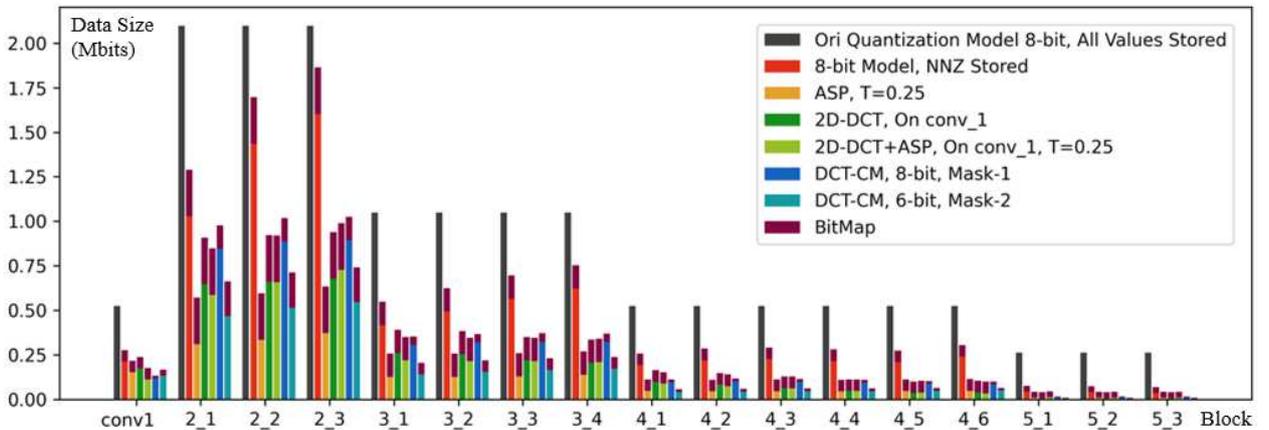
Fig. 1. By using different sparsity-based compression algorithms, the Number of Non-Zero (NNZ) values of feature map in each block of ResNet-50. T for Threshold in ASP. Mask-1/2 for different typical mask settings in DCT-CM, which are selected by experiments.

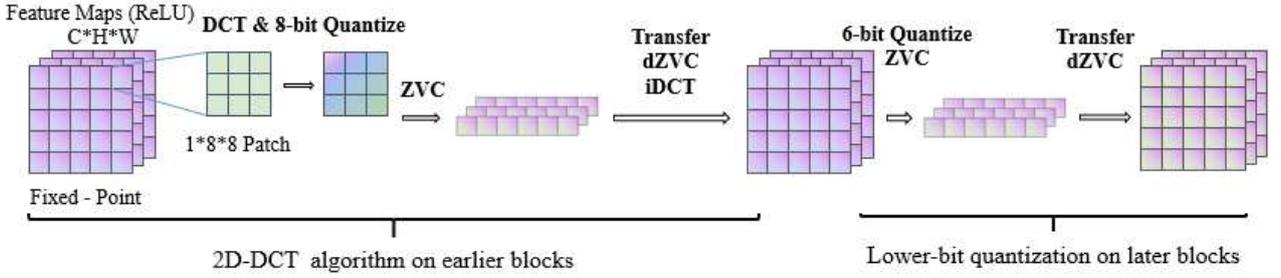

Fig. 2. 2D-DCT: Different compression strategies on earlier or later blocks in CNN

then we can improve the sparsity of activation. As a result, the DCT-CM will restructure the filters on CNN, then this algorithm has to be used at the start of training instead of fine-tuning.

The 2D-DCT has two disadvantages in the inference that it cannot be deployed on small-scale feature maps and it only shows a high compression ratio on earlier blocks in image processing tasks. In contrast, DCT-CM can be used on any activation, which brings much more hardware reusability. Furthermore, the shape (or the length) of masks can be adjusted, e.g. from 8 to 16 or 4. The selection depends on the balance of computation cost and compression ratio. Besides, the valid region of mask groups can be also set by NAS.

By using the high sparsity of feature maps, [33] chooses feature selection on high dimensional vectors and [18, 29] change the structure of CNNs. On the contrary, the DCT-CM transforms the feature map on the channel dimension instead of the special dimension. Therefore, this algorithm reorders the channels of activations during training and provides static sparsity for the compression module.

Considering the computational cost, compared with 2D-DCT, decompression modules of DCT-CM will read fewer values, since parts of the data are set to zero by the mask. Thus, the DCT-CM which saves low-frequency information regularly can reduce both the bandwidth and the inverse-transform computational cost. DCT-CM operates every single pixel on the channel dimension of feature maps and uses fewer matrix operations for each unit (4), (5).

Furthermore, the transformation matrix $\mathbf{A}$ in (5) can be fused into next convolutional layers by the linear transformation on weights. Then, we use the fact that ZVC describes both the compression algorithm and the data flow of compressed feature map in convolutional filters. As a result, we can cut off the extra computational budget caused by IDCT. E.g., we take a transformed slice, $\mathbf{Y}_{\text{freq-domain}} \in \mathbb{R}^{8 \times 1 \times 1}$, and use (6) to get $\mathbf{Y}_{\text{reshape}}$. A block of weights in convolutional kernels is $\mathbf{W} \in \mathbb{R}^{C_{out} \times 8}$, $\mathbf{Y}_{\text{out}}$ is calculate by (7). Then we only need to save $\mathbf{W}^*$ of (8) for the inference and reduce inverse-transform computational budget.

$$\mathbf{Y}_{\text{reshape}} = \mathbf{A}^T \mathbf{Y}_{\text{freq-domain}} \quad (6)$$

$$\mathbf{Y}_{\text{out}} = \mathbf{W} \mathbf{Y}_{\text{reshape}} \quad (7)$$

$$\mathbf{W}^* = \mathbf{W} \mathbf{A}^T \quad (8)$$

Therefore, DCT-CM can provide more flexibility in convolutional architecture design and higher throughput than 2D-DCT. Furthermore, the DCT-CM algorithm introduces a little computation of matrix multiplication compared with the convolution. E.g., for each block and its outputs in c4 in ResNet-50 [4], 0.2G MACs (Multiply Accumulates) is needed by convolutional layers to compute the output feature maps, but DCT-CM only need 0.4M MACs (0.2%) with the mask type 2/8 after applying the further optimization that skips the zero computation caused by the mask.

### C. Approximate Sparsity Preprocess and L1-Loss

Feature maps can be sparse after extracted from ReLU in neural networks. Furthermore, small values in these feature maps can be dropped to increase the sparsity immediately. It is noticeable that this method is different from lower-bit quantization because the precision of large values is still kept the same as the original quantization.

The proposed preprocessing algorithm, ASP, (9) is very simple, which does not need to constrain the data reading. The parameters of the new model using this method can be set by fine-tuning.

$$\textit{Act\_value} = 0 \text{ if } \textit{Act\_value} < \textit{Threshold} \text{ else } \textit{Act\_value} \quad (9)$$

Although ASP is a lossy algorithm, it keeps good accuracy. This straightforward sparsity algorithm runs without a homologous decoder. Thus, we insert ASP into 2D-DCT and DCT-CM to further improve the sparsity of feature maps.

Another widely-used way to increase the dynamic sparsity of feature maps is to regularize activations by extra loss functions, such as L0, L1, or smooth L1. The effect of L1 loss in our work will be discussed in Section IV.

## IV. RESULTS

### A. Experimental Setup

We train an 8-bit quantization ResNet-50 [4] as a baseline on CIFAR-10 [34]. Because the proposed algorithms are sparsity-based, we calculate the compression ratio by ZVC encoder. All the feature maps are extracted after the last ReLU activations of each block in ResNets.

### B. Selecting Parameters and Combinations

During the experiments, we aim at achieving higher compression ratio as much as possible within 0.5% accuracy drop in image classification. Table I. shows that the proposed

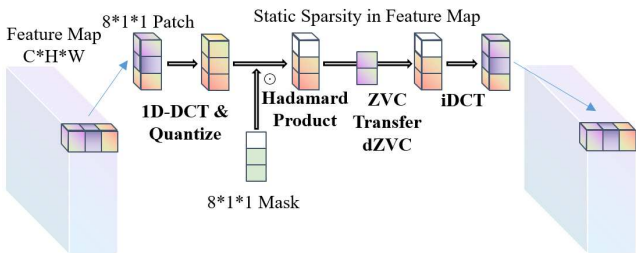

Fig. 3. DCT-CM uses DCT and mask to improve sparsity during transfer.

DCT-CM can express a higher compression ratio of 2.9× with less accuracy dropped of 0.39%. Table II. lists representative settings and the results of our experiments. When deploying 2D-DCT algorithms, we apply 2D-DCT to the first layer (stage c1 in ResNet-50, represented by parameter '1') or earlier blocks (stage c1 and c2 in ResNet-50, represented by '1 & 2 blocks') but apply low-bit quantization to the rest. Besides, 2D-DCT performs well on accuracy rather than compression ratio, which implies it can be used on complex tasks. The L1 loss cannot improve the sparsity further based on proposed methods.

Though ASP shows high sparsity and compression ratio, it works worse with a higher threshold. Therefore, we infer that ASP is not qualified for the model quantized in lower-bit but can be used as an optional plug-in with transform-based algorithms. Differently, the proposed novel method, DCT-CM, can express both high validation accuracy and high compression ratio in a variety of parameters with appropriate masks (M-1 and M-2 in Table II. for different mask settings, which are the same as them showed in Fig. 1. M1: [4, 6, 4, 2, 1] / 8, M2: [2, 4, 3, 2, 1] / 8, which represent [the amount of low-frequency information to save in different stages] / length of mask). Our experiments indicate the flexibility of the parameter selection of DCT-CM, which implies its prospect in complicated tasks. Besides, by combining higher-bit and ASP, DCT-CM can improve the compression ratio and also maintain accuracy.

## V. CONCLUSION & FUTURE WORK

In this paper, we propose novel transform-based and sparsity-based algorithms DCT-CM to reduce the data transfer bandwidth and power consumption for CNN inference. By combining it with ZVC, we achieve a compression ratio of 2.9× with negligible accuracy dropped on a CNN model with an 8-bit fixed-point quantization scheme for inference. Designing a hardware accelerator architecture for DCT-CM and combining NAS into masks selection are our future directions.

TABLE I. INFLUENCE OF DIFFERENT PARAMETERS

| Algorithm | Parameter | Drop (%) | C. R. |
|---|---|---|---|
| Low-bit | 8-bit (Baseline) | 0 | 1.5× |
| | 5-bit | -0.77 | 2.2× |
| DCT-2D | 1 | +0.05 | 2.5× |
| | **1 0.5ASP** | **-0.48** | **2.6×** |
| | 1 & 2 blocks | +0.02 | 2.4× |
| ASP | **0.25** | **-0.25** | **3.9×** |
| | 0.25 L1 | -0.33 | 3.9× |
| | *0.5* | *-1.97* | *4.4×* |
| DCT-CM | 10-bit M-1 | -0.43 | 2.2× |
| | *10-bit M-1 0.125* | *-0.87* | *3.1×* |
| | **8-bit M-1** | **-0.39** | **2.9×** |
| | *6-bit M-2* | *-0.71* | *4.4×* |

TABLE II. COMPRESSION RATIO (C.R.) OF DIFFERENT PROPOSED ALGORITHMS WITH ZVC (RESNET-50, CIFAR-10)

| Algorithm | Acc. or Drop (%) | C.R. |
|---|---|---|
| Baseline | 95.08 | 1.5× |
| cDMA [18] | - | 1.1× |
| *EBPC [10] (ResNet-34)* | - | *2.45×* |
| JPEG-ACT [16] | -0.68 | 1.9× |
| ASP (Our) | -0.25 | 3.9× |
| DCT-2D (Our) | +0.05 | 2.5× |
| **DCT-CM (Our)** | **-0.39** | **2.9×** |

REFERENCE

[1] A. Krizhevsky, I. Sutskever, and G. E. J. C. o. t. A. Hinton, "Imagenet classification with deep convolutional neural networks," in Communications of the ACM, vol. 60, no. 6, pp. 84-90, 2017.

[2] R. Girshick, "Fast r-cnn," in Proceedings of the IEEE international conference on computer vision, 2015, pp. 1440-1448.

[3] T. Young, D. Hazarika, S. Poria, and E. J. i. C. i. m. Cambria, "Recent trends in deep learning based natural language processing," in ieee Computational intelligenCe magazine, vol. 13, no. 3, pp. 55-75, 2018.

[4] K. He, X. Zhang, S. Ren, and J. Sun, "Deep residual learning for image recognition," in Proceedings of the IEEE conference on computer vision and pattern recognition, 2016, pp. 770-778.

[5] K. Simonyan and A. J. a. p. a. Zisserman, "Very deep convolutional networks for large-scale image recognition," 2014.

[6] M. Sandler, A. Howard, M. Zhu, A. Zhmoginov, and L.-C. Chen, "Mobilenetv2: Inverted residuals and linear bottlenecks," in Proceedings of the IEEE conference on computer vision and pattern recognition, 2018, pp. 4510-4520.

[7] J.-H. Luo, J. Wu, and W. Lin, "Thinet: A filter level pruning method for deep neural network compression," in Proceedings of the IEEE international conference on computer vision, 2017, pp. 5058-5066.

[8] B. Zoph and Q. V. J. a. p. a. Le, "Neural architecture search with reinforcement learning," 2016.

[9] M. Tan and Q. Le, "EfficientNet: Rethinking Model Scaling for Convolutional Neural Networks," in International Conference on Machine Learning, 2019, pp. 6105-6114.

[10] L. Cavigelli and L. Benini, "Extended bit-plane compression for convolutional neural network accelerators," in 2019 IEEE International Conference on Artificial Intelligence Circuits and Systems (AICAS), 2019, pp. 279-283: IEEE.

[11] J. Albericio, P. Judd, T. Hetherington, T. Aamodt, N. E. Jerger, and A. J. A. S. C. A. N. Moshovos, "Cnvlutin: Ineffectual-neuron-free deep neural network computing," in ACM SIGARCH Computer Architecture News, vol. 44, no. 3, pp. 1-13, 2016.

[12] A. Parashar et al., "Scnn: An accelerator for compressed-sparse convolutional neural networks," in ACM SIGARCH Computer Architecture News, vol. 45, no. 2, pp. 27-40, 2017.

[13] Y.-H. Chen, J. Emer, and V. J. A. S. C. A. N. Sze, "Eyeriss: A spatial architecture for energy-efficient dataflow for convolutional neural networks," in ACM SIGARCH Computer Architecture News, vol. 44, no. 3, pp. 367-379, 2016.

[14] S. Han et al., "EIE: efficient inference engine on compressed deep neural network," in ACM SIGARCH Computer Architecture News, vol. 44, no. 3, pp. 243-254, 2016.

[15] S. Zhang et al., "Cambricon-x: An accelerator for sparse neural networks," in 2016 49th Annual IEEE/ACM International Symposium on Microarchitecture (MICRO), 2016, pp. 1-12: IEEE.

[16] R. D. Evans, L. Liu, and T. M. Aamodt, "Jpeg-act: accelerating deep learning via transform-based lossy compression," in 2020 ACM/IEEE 47th Annual International Symposium on Computer Architecture (ISCA), 2020, pp. 860-873: IEEE.

[17] M. Rhu, N. Gimelshein, J. Clemons, A. Zulfiqar, and S. W. Keckler, "vDNN: Virtualized deep neural networks for scalable, memory-efficient neural network design," in 2016 49th Annual IEEE/ACM International Symposium on Microarchitecture (MICRO), 2016, pp. 1-13: IEEE.

[18] M. Rhu, M. O'Connor, N. Chatterjee, J. Pool, Y. Kwon, and S. W. Keckler, "Compressing DMA engine: Leveraging activation sparsity for training deep neural networks," in 2018 IEEE International Symposium on High Performance Computer Architecture (HPCA), 2018, pp. 78-91: IEEE.

[19] A. Jain, A. Phanishayee, J. Mars, L. Tang, and G. Pekhimenko, "Gist: Efficient data encoding for deep neural network training," in 2018 ACM/IEEE 45th Annual International Symposium on Computer Architecture (ISCA), 2018, pp. 776-789: IEEE.

[20] Q. Xiao, Y. Liang, L. Lu, S. Yan, and Y.-W. Tai, "Exploring heterogeneous algorithms for accelerating deep convolutional neural networks on FPGAs," in Proceedings of the 54th Annual Design Automation Conference 2017, 2017, pp. 1-6.

[21] A. Tumeo, M. Monchiero, G. Palermo, F. Ferrandi, and D. Sciuto, "A pipelined fast 2D-DCT accelerator for FPGA-based SoCs," in IEEE Computer Society Annual Symposium on VLSI (ISVLSI'07), 2007, pp. 331-336: IEEE.

[22] J. Huang, M. Parris, J. Lee, and R. F. J. A. T. o. E. C. S. Demara, "Scalable FPGA-based architecture for DCT computation using dynamic partial reconfiguration," vol. 9, no. 1, pp. 1-18, 2009.

[23] H. El-Banna, A. A. El-Fattah, and W. Fakhr, "An efficient implementation of the 1D DCT using FPGA technology," in Proceedings of the 12th IEEE International Conference on Fuzzy Systems (Cat. No. 03CH37442), 2003, pp. 278-281: IEEE.

[24] C. Loeffler, A. Ligtenberg, and G. S. Moschytz, "Practical fast 1-D DCT algorithms with 11 multiplications," in International Conference on Acoustics, Speech, and Signal Processing, 1989, pp. 988-991: IEEE.

[25] B. W. Bader and T. G. J. S. J. o. S. C. Kolda, "Efficient MATLAB computations with sparse and factored tensors," vol. 30, no. 1, pp. 205-231, 2008.

[26] A. Moffat and J. Zobel, "Parameterised compression for sparse bitmaps," in Proceedings of the 15th annual international ACM SIGIR conference on Research and development in information retrieval, 1992, pp. 274-285.

[27] E. L. Hauck, "Data compression using run length encoding and statistical encoding," ed., 1986.

[28] X. Zeng et al., "Addressing Irregularity in Sparse Neural Networks through a Cooperative Software/Hardware Approach," in IEEE Transactions on Computers, 2020.

[29] D. Gudovskiy, A. Hodgkinson, and L. Rigazio, "DNN feature map compression using learned representation over GF (2)," in Proceedings of the European Conference on Computer Vision (ECCV), 2018, pp. 0-0.

[30] A. Aimar et al., "Nullhop: A flexible convolutional neural network accelerator based on sparse representations of feature maps," in IEEE transactions on neural networks and learning systems, vol. 30, no. 3, pp. 644-656, 2018.

[31] G. K. J. I. t. o. c. e. Wallace, "The JPEG still picture compression standard," in IEEE transactions on consumer electronics, vol. 38, no. 1, pp. xviii-xxxiv, 1992.

[32] I. J. Goodfellow, J. Shlens, and C. J. a. p. a. Szegedy, "Explaining and harnessing adversarial examples," in arXiv preprint arXiv:1412.6572, 2014.

[33] Y. Zhang, J. Wu, and J. Cai, "Compact representation for image classification: To choose or to compress?," in Proceedings of the IEEE Conference on Computer Vision and Pattern Recognition, 2014, pp. 907-914.

[34] Saito, "chainer-cifar10: Various CNN models including for CIFAR10 with Chainer," 2018, original-date: 2015-06-09T14:39:43Z. [Online]. Available: https://github.com/mitmul/chainer-cifar10